\documentclass{aa}            
\usepackage{graphicx}
\usepackage{txfonts}
\begin{document}
\title{The effect of heat conduction on the interaction of disk and
corona around black holes}
\author{E. Meyer-Hofmeister and F. Meyer}
\offprints{Emmi Meyer-Hofmeister; emm@mpa-garching.mpg.de}
\institute{
Max-Planck-Institut f\"ur Astrophysik, Karl-
Schwarzschildstr.~1, D-85740 Garching, Germany
} 

\date{Received: / Accepted:}

\abstract{Heat conduction plays an important role in the balance
between heating and cooling in many astrophysical objects,
e.g. cooling flows in clusters of galaxies. 
Here we investigate the effect of heat conduction on the interaction 
between a cool disk and a hot corona around black holes.  Using the 
one-radial-zone approximation, we study the vertical structure of the 
disk corona and derive evaporation and coronal mass flow rates for 
various reduced thermal conductivities. We find lower evaporation 
rates and a shift in the evaporation maxima to smaller radii.
This implies that the spectral state transition occurs at a lower mass 
flow rate 
and a disk truncation closer to the black hole. Reductions of thermal
conductivity are thought to be magnetically caused and might vary from
object to object by a different configuration of the magnetic fields.

} 
\titlerunning {Effects of reduced heat conduction}
\maketitle

\section{Introduction}

Heating by thermal conduction is an important process in many
astrophysical objects. The balance between cooling and
heating processes determines the structure of hot matter in contact 
with cooler regions. For the case of a dwarf nova accretion
disk, we investigated this interaction (Meyer \& Meyer-Hofmeister 1994)
and found that a ``siphon flow'' leads to evaporation of the
disk. The same happens in disks around neutron stars and in both galactic and 
supermassive black holes (Meyer-Hofmeister \& Meyer 1999, Meyer et
al. 2000). The heat conduction is an essential element; otherwise the 
coronal gas would not lose energy, but would instead become hotter and reach 
virial temperatures as it is present in advection-dominated flows. 

Recently the same physical process was discussed by several authors
in connection with the cooling flow problem for the hot intra-cluster
medium (ICM). 
The work of Medveden \& Narayan (2001) focuses on the question as to
what degree chaotic magnetic fields suppress conduction relative to the
Spitzer level. They find that thermal conduction in a weakly
collisional plasma with turbulent magnetic fields approaches the
Spitzer limit. Zakamska \& Narayan (2003) find from the
investigation of five galaxy clusters that the Spitzer formal 
with a conduction coefficient reduced to about 30\% gives a good 
description of the observed radial profiles of electron density and 
temperature.
Ghizzardi et al. (2004) discuss which fraction of the
Spitzer value would be appropriate xplaining the Virgo/M87
observations.  Voigt \& Fabian (2004), on the other hand, found
support for an unhindered heat conduction from their 
analysis of 16 galaxy clusters using Chandra data. One third of the 
Spitzer value was used 
in hydrodynamic cosmological simulations of galaxy clusters by 
Dolag et al. (2004). In this context Soker et al. 2004 suggest a heat 
conduction along magnetic field lines.  Okabe
\& Hattori (2004) suggest a suppression of heat conduction by
magnetic fields
generated most strongly in the direction perpendicular to the
temperature gradient. All these results point to a reduced heat
conduction in many sources.

In the original context of disk/corona interaction, the effect of
reduced thermal conduction has already been considered by a scaling
procedure (Meyer et al.2000). In the present work we evaluate the 
effect of heat conduction on the evaporation of accretion disks in
detail. Connected with the evaporation
efficiency, this means a possible change in the resulting
truncation radius. Motivation also comes from the fact that new
computations including the irradiation of the coronal gas from the
inner region (Meyer-Hofmeister et al. 2005 (hereafter MLM05), Liu et 
al. 2005) have 
led to radii that seem larger than indicated by observations (Yuan \& 
Narayan 2004). In these works the thermal conduction
was taken according to the standard value derived by Spitzer 
(1962). The question then arises whether a reduced heat conduction could be 
present in the disks around black holes. Here we study the effect of heat 
conduction on evaporation and the truncation of the accretion disk.

In Sect. 2 we give a short description of the accretion geometry and 
the interaction of corona and disk. In Sect. 3 we 
present the results for reduced heat conduction, and a discussion of the
consequences and conclusions follow in Sect. 4.

\section{The physics of a hot corona above a cool disk}
In the work by Meyer et al. (2000), the interaction of disk and corona 
was approximated by a one-zone model 
that allows us to evaluate the
evaporation of mass from the cool disk into a coronal flow above the
disk. The basic process is the following. The coronal temperature is
kept up by friction that releases gravitational energy from
accretion. The temperature gradient between the corona and cool disk
causes a heat flow. The density in the corona adjusts itself by
evaporation of gas from the disk so that this heat flow is balanced by
evaporation, radiation, and advection. This equilibrium density in the
corona implies a coronal mass accretion rate that is fed by gas from
the disk and determines the disk evaporation rate.  

The five equations describing the process are: 
(1) the equation of continuity, (2) the
$z$-component of momentum equation, (3) and (4) the two energy equations
for ions and electrons, and (5) the equation for the thermal conduction 
for a fully ionized plasma (see Liu et al. 2002). The effect of
Compton cooling and heating of coronal electrons by photons from the 
central area was recently worked out in more detail and found to be 
important for the evaporation process (MLM05).
The five dependent variables are pressure, ion and electron temperature,
vertical mass flow, and heat flux. The boundary conditions are 
taken (a) at the bottom of the corona, temperatures given, no heat inflow,
and (b) at some height (free boundary), sound transition, no influx of
heat. 

The general picture is as follows. Controlled by the interaction 
of the disk and corona, gas evaporates from
the disk into the corona and flows inward in the form of a hot
advection-dominated flow. The mass flow rate in the thin disk 
is thereby diminished. The evaporation rate increases with decreasing distance 
from the compact object, but reaches a maximum at a certain distance 
$R_{\rm{evmax}}$ from the black hole. The balance between this maximum 
rate and the mass flow rate in the disk determines the mode of
accretion in the inner region. Only if the mass flow rate in the
outer disk is higher than this maximum value of the evaporation rate,
the disk ``survives'' this reduction in mass flow and continues
inward. Otherwise, for lower mass flow rates in the disk,
the disk becomes truncated at a certain distance (where all mass is
evaporated). From then on all mass flows inward via the coronal/ADAF
flow. This is the same picture as the configuration of accretion flow
in different spectral states derived by Esin et al (1997) in
connection with the application of the scheme of advection-dominated 
accretion to Nova Muscae 1991.  

The agreement between the results of the one-zone approximation and 
observed features seems to confirm that our model describes the 
qualitative picture of the change from accretion via a disk to the 
advection-dominated flow (ADAF). Such agreement was found 
for X-ray transient sources in connection with the spectral
transitions (Meyer et al. 2000), especially the hysteresis in the 
transition luminosity (MLM05), as well as accretion disk 
evolution (Meyer-Hofmeister \& Meyer 1999), and also in connection
with the truncated disks in low-luminosity AGN (Liu \& Meyer-Hofmeister
2001). But if we want to extract detailed
qualitative results, we have to keep in mind that a number of
simplifications used in the description lead to uncertainties of the
computational results. Two free parameters (for given mass of the 
central object and chemical abundance) enter into the evaluation of the
evaporation rate: For the viscosity parameter $\alpha$, we use the value 0.3
suggested by observations (for a discussion see MLM05). Note that in 
our one-zone calculations, $\alpha$
parameterizes frictional heat release and also radial flow of mass and
angular momentum. Its value is not directly comparable to the value
of the standard Shakura and Sunyaev $\alpha$ value (Shakura \&
Sunyaev 1973). For a comparison with the ratio of stress to pressure in
magneto-hydrodynamic calculations, see a discussion by 
Hawley \& Krolik (2001). The second parameter in the evaluation of the
evaporation rate is the thermal conduction.

From our recent work on hysteresis in the spectral state transitions, it
became obvious that the irradiation of the corona has an important
effect on evaporation rates and truncation of the cool accretion disk.
We denote by $R_{\rm{evmax}}$ the shortest distance at which truncation 
still can occur. Earlier computations without including the 
irradiation (Meyer et al. 2000) had led to a distance of a few hundred 
Schwarzschild radii. Including the irradiation of the corona, we now obtain
values about 2.5 times larger. This is the innermost disk truncation; 
for lower accretion rates the truncation lies farther outward, while for
higher rates the disk is not truncated anymore. How do these radii 
compare with radii found from observations? The observed radii should 
be equal or larger than $R_{\rm{evmax}}$, depending on the mass flow
rate in the disk. Despite the fact that one should be careful with
such a comparison and not take signatures of a reflecting component as
the disk truncation, the observations point to lower values.  
A compilation of radii in the paper of 
Yuan \& Narayan (2004, Fig.3) suggests values around 100 Schwarzschild 
radii. Esin et al. (2001) use the multiwavelength observations of the
X-ray nova XTE J1118+480 to constrain the accretion geometry and find
a disk truncation in outburst at $\ge$ 55 Schwarzschild radii.
The question arises whether a reduced thermal conduction could be
present in the accretion disks around black holes and how this would
change the evaporation process. In the next section we show our 
results for reduced heat conduction.

\section{Computational results}

We take a black hole mass of $M=6M_\odot$. Results can be scaled for 
other masses. We, therefore, show the evaporation rates
measured in Eddington accretion rate 
$\dot M_{\rm{Edd}}=L_{\rm{Edd}}/0.1c^2$ with 
$L_{\rm{Edd}}=4 \pi GMc/\kappa_{\rm{es}}$, $\kappa_{\rm{es}}$ 
electron scattering opacity. The distances are measured in
Schwarzschild radii $R_{\rm S} =2GM/ c^2$. 
We use the computer code as described in recent work (LMM05). 
The heat conduction coefficient $\kappa_0$ 
\begin{equation}
\kappa_0=10^{-6} \frac{\rm{g cm}}{\rm{s^3 K^{7/2}}}
\end{equation}
enters the equation for the heat flux  
\begin{equation}
F_c=-\kappa_0 T^{5/2}dT/dz
\end{equation}
with T temperature, and $z$ height above midplane. 
For our calculations we take a fraction $\lambda$ 
of the standard value for a fully ionized plasma (Spitzer 1962)

The heat conduction also enters in our lower boundary condition at the
bottom of the corona, and, without going into detail, this results in
a change of the relation between pressure and heat flux (Meyer et
al. 2000, Eq.54) to $P=A \cdot F_c \lambda ^{-1/2}$ (compare also 
F.K. Liu et al. 1995). 

For taking into account Compton cooling and heating by radiation from 
the innermost region we use $L=\eta\dot M c^2$, 
$\eta=0.1$ for the luminosity of the
central source for various heights of the corona along with  
$\eta=0.1$ with $\dot M$ central mass accretion rate.
We calculate the irradiation flux from the central region as 
in recent work (MLM05, Liu et al. 2005).

\subsection{The evaluation of evaporation rates consistent with
irradiation from the central region}
We calculate the evaporation rates for a series of distances from the
black hole. We consider both cases of irradiation of the corona: (1)
hard radiation from an inner region filled with a hot coronal gas 
(advection-dominated accretion flow, ADAF, or (2) soft radiation from
an accretion disk in the inner region. For our analysis of the hard
state we take a mean photon energy of 100 keV for the radiation from
the central source. We are mainly interested in
case (1) since this describes the situation when the disk is truncated. 
The truncation radius depends on the mass flow rate in the cool disk.
This is the situation as long as the system is in the hard state.
Case (2) is relevant here only at the moment when the mass flow rate 
decreases from a rate above the maximal evaporation rate (the disk
then extends inward to the last stable orbit) to a rate below the
maximal evaporation rate, and the disk breaks up at the distance 
$R_{\rm{evmax}}$.

The evaporation rate is a function of the distance and of the
radiation from the central region, if
the effect of irradiation is included. In the one-zone approximation,
the mass flow entering 
and leaving is modeled appropriately for a corona at the inner disk
edge, i.e. the mass flow rate in the disk equals the evaporation rate.
As this mass flow continues into the central region it causes the radiation
which then irradiates the corona at the truncation radius. 
In other words, the adequate irradiation to be included corresponds
to the evaporation rate itself. Therefore an iterative procedure is 
necessary to determine the evaporation  rate consistent with the
irradiation; evaporation rates are evaluated for different values of 
radiation from the inner region until we find the evaporation rate
consistent with the irradiation produced by exactly a mass flow rate
equal to the evaporation rate.

In recent work we included the irradiation effect  
(MLM05, Liu et al. 2005) only for the maximal evaporation rates at 
$R_{\rm{evmax}}$, which is important for the spectral transition. For the
 present investigation we now have determined the 
rates all all distances consistent with the irradiation effect. Figure
1 thus displays a sequence of disk truncation radii. These
results then also allow us to discuss the slope of the evaporation rate
- distance relation.

In Fig. 1 we also show our new results for the reduced heat conduction. We 
evaluated evaporation rates for the standard 
Spitzer value heat conduction and a reduction to 50\% and 20\% of this value.
As already pointed out in the earlier work, in the hard state the 
electron energy decreases with increasing distance, the Compton
cooling from the central source turns into heating, and the irradiation
yields higher evaporation rates than without irradiation.

\begin{figure}
\centering
\includegraphics[width=8.3cm]{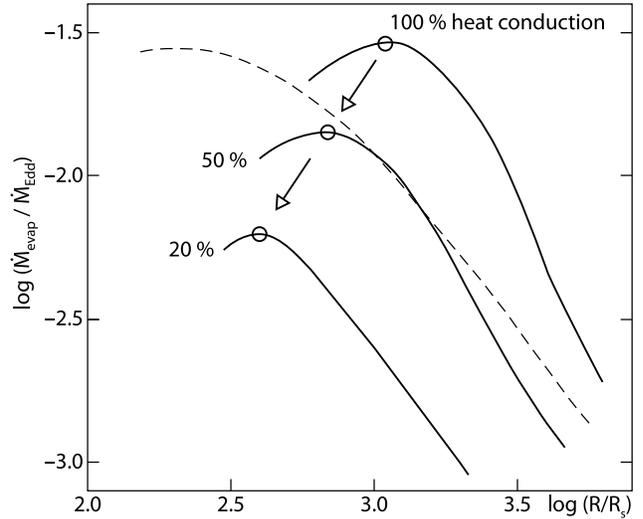}
\caption[.]{Effect of reduced heat conduction: Evaporation rate as a function
of distance from the black hole (hard spectral state). Solid
lines:  Thermal conduction according to standard Spitzer rate and 
reduced to 50\% and 20\%. 
Dashed line: 100\% heat conduction, but without irradiation. Reduced heat 
conduction leads to disk truncation farther 
inward. The maxima give locations of innermost disk truncation at 
spectral transition}
\end{figure}

\subsection{The evaporation rates for hard irradiation}
As shown in Fig.1, reduced heat conduction strongly affects both the
evaporation and the location of the maxima. For example if one assumed 
a mass flow of 0.005 $\dot M_{\rm{Edd}}$ from the outer
disk with 100\% of the Spitzer value of heat conduction, the disk would be 
truncated at about 3850 Schwarzschild radii. But with a heat conduction
reduced to 50 and 20\%, respectively, the truncation moves in to 1860
and 570 $R_{\rm S}$, a factor of almost 7 closer to the center in the 
latter case. For low mass accretion rates 
when the disk truncation is at large distances, the irradiation effect
becomes small. In Fig.1 we show for comparison results for 100\%
heat conduction without irradiation (MLM05). It can be seen that the 
curves for 100\%
heat conduction with and without irradiation begin to converge  
at large distances.

Interesting is the shift of the evaporation maxima in distance,
from about 1100 $R_{\rm S}$ to 700 and 400 $R_{\rm S}$ for 
reduced heat conduction. Also the amplitude of the maxima is lowered. 
For a heat conduction reduced to 20\% of the standard Spitzer value, 
we find an innermost disk truncation at $R_{\rm S}$ corresponding to a 
mass flow rate of 0.006 $\dot M_{\rm{Edd}}$. 
This rate can be compared with the results of Maccarone (2003), who
critically analyzes observations to deduce the luminosity of neutron
star and black hole transient binaries at the soft/hard spectral 
transition. Our value derived for reduced heat conduction is 
relatively low, but lies in the range found from the observations.

\subsection{The coronal structure for hard irradiation}
Reduced heat conduction changes the vertical structure of the corona.
We compare the structure at peak evaporation
rate for the two cases: heat conduction either equal to the standard Spitzer
value or reduced to 20 \% (compare Fig.1). In Fig.2 we show the
electron and ion temperature as a function of $z/R$. Temperatures
are measured in virial temperature 
$T_{\rm{vir}}=GMR \cdot  \mu/ \Re$.  
There is a clear change with heat conduction. For the reduced value,
the electron temperature is lower and the ion temperature is higher 
than for the more effective heat conduction. 
Note that the virial 
temperatures depend on the distance from the black hole, which is
different in the two cases. 
The corresponding values of pressure at the lower
boundary of the corona are log P =7.04 
$ \rm{gcm^{-1}sec^{-2}}$ (100\% heat conduction) and 7.53
$ \rm{gcm^{-1}sec^{-2}}$ (20\%), which is a factor of 3 higher in the 
latter case.

\begin{figure}
\centering
\includegraphics[width=7.0cm]{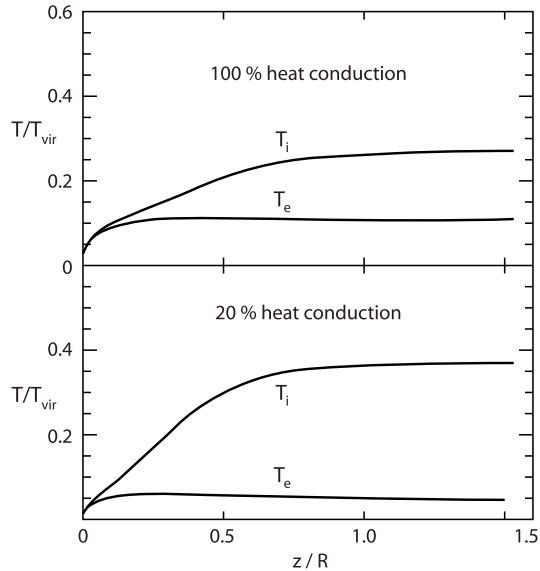}
\caption[.]{Coronal temperatures:
Electron and ion temperatures (measured in virial temperature) at
height $z/R$, $R$ distance from black hole, $z$ height above midplane.
}
\end{figure}

\subsection {Computational results for soft irradiation}
The effect of soft irradiation is only of interest here
at the moment the disk becomes truncated by evaporation, i.e. the
soft/hard transition. Compton cooling leads to lower evaporation rates (MLM05).
Including reduced heat conduction decreases the evaporation
rates even more. In our computations we encountered problems finding 
solutions for the coronal structure to fulfill
our upper boundary conditions such as sound transition and no heat inflow 
from infinity. Further work is needed to clarify whether the condition
$P \rightarrow 0 $ for large $z$ can be achieved with a subsonic structure.
                                                                            
For heat conduction reduced by a factor of 5 and soft 
irradiation, we found 
the maximal evaporation rate of $ \dot M_{\rm{evap}}$/ $\dot
M_{\rm{Edd}}$=0.004 at about 400 Schwarzschild radii. The
difference between the rates for hard and soft irradiation
allows us to understand the hysteresis in the transition luminosity of
X-ray transients. For reduced heat conduction the difference becomes
small, less than a factor of 2, and much less than what was found for
unreduced heat conduction. This might point to a problem understanding
the hysteresis seen in X-ray nova outburst light curves for the case 
of strongly reduced heat conduction; if not other effects enlarge this 
difference. Such further effects might be due to time delays in the 
transition and recondensation of hot coronal gas into the disk. 
We are presently investigating this situation (Liu \& Meyer).

\section{Discussion}
\subsection{Disk truncation radius as a function of the mass accretion
rate}
The curves in Fig.1 show both the dependence of evaporation rates on the
distance and the disk truncation for a given mass flow rate in the
disk. It might be interesting to analyze observations of X-ray
transients to see how the inner disk edge moves inward or outward 
as a result of increasing or decreasing mass flow.

Our theoretical analysis only allows us to make rough predictions due to
the uncertainties in the approximations. The curves in Fig.1 are not
straight lines. For standard heat
conduction and for a range of distances of $R_{\rm{tr}}/R_{S}$
from $10^{3.3}$ to $10^{3.9} \rm{cm}$ we find a slope of

\begin{equation}
\Delta \rm{log}R_{\rm{tr}}/ \Delta \rm{log} \dot M \approx -0.5.
\end{equation}
We find a slightly steeper slope of about -0.8 
for a reduction to 20\% (for distances $R_{\rm{tr}}/R_{S}$ from 
$10^{2.8}$ to $10^{3.3}\rm{cm}$).

If we compare our results with changes of the inner
disk location of the X-ray transient source XTE J1118, in particular
we find for truncation radii and luminosities the following values
 in the literature: in outburst 
log $R_{\rm{tr}}$/log$R_{\rm S}$=1.74 for log $L/L_{\rm{Edd}}$=-1.7,
(Esin et al.2001) and in quiescence 
log $R_{\rm{tr}}$/log$R_{\rm S}\approx 4$ for log $L/L_{\rm{Edd}}$=-8.5    
(McClintock at al. 2003). Note that the source is  also in the hard 
spectral state in outburst. This gives the value 0.33 for  
$\Delta \rm{log}R_{{tr}}/ \Delta \rm{log} \dot L$.
If we want to translate this into a radius-mass accretion rate relation, 
we need to keep in mind that for low mass accretion rates, radiation processes 
scale with particle density squared, thus leading roughly to a scaling
of the luminosity with $\dot M^2$. This would yield an approximate 
agreement between the theoretical model and observations.

A relation between truncation radius and luminosity for various sources from 
bright AGN and X-ray binaries to dim objects as Sgr A* is shown in
the investigation of X-ray bright, optically normal galaxies by
Yuan \& Narayan (2004, Fig.3)

\subsection {Is heat conduction always the same?}
A number of discussions have arisen around cooling flows and
magnetic fields. We note that in the more general frame of
condensation and evaporation between corona and cool disk evaporation,
flows can straighten out fields from the cool gas as they carry the
fields with them. It seems conceivable that the value of the thermal
conduction varies from object to object and also in one object with
time depending on the accretion flow history.
Zdziarski et al. (2004) study the long-term variability of the X-ray 
nova GX 339-4 and discuss the spectral changes in different
outbursts. Especially during the two recent outbursts, the hard/soft 
transition in the second one occurred at a  higher luminosity than that
in the preceding one. They argued that this
difference could arise from the different accretion disk history. The 
magnetic field pattern in the disk might be influenced by this history
of more or less mass flow in the disk.

\section{Conclusions}
We have evaluated how a reduced heat conduction affects the 
evaporation of gas from a cool disk to a hot coronal flow/ADAF.
The physical situation of evaporation (or condensation) between
hot (``coronal'') and cool (``disk'') gas is the same around stellar
black holes and in galaxies and clusters of galaxies. For clusters 
of galaxies, reduced heat conduction seems to be 
supported by observations in several cases.

In our theoretical modeling, evaporation rates become
lower with reduced heat conduction, and the location where the
evaporation efficiency reaches its 
maximum moves inward by a factor of 7 for the reduction to 20\%. 
This moves the truncation radii closer to an agreement with
observations, but a significant difference exists still. 
On the other hand, the change of truncation radii with accretion rate 
$\Delta \log R_{\rm{tr}}/ \Delta \log \dot M$ in our results is in 
reasonable agreement with the numbers derived from observations for 
galactic and supermassive black holes.

Interesting is the strong dependence of spectral transition on heat 
conduction. A further reduction might arise from a different
magnetic field situation. This can affect the transition from a very
bright state to a very dim state in the AGN of elliptical galaxies as 
suggested by Churasov et al. (2005).

\end{document}